\documentclass[amsmath,pre,aps,groupedaddress,twocolumn]{revtex4}
\usepackage{graphicx}

\newcommand{\diff}[2]{\frac{d{#1}}{d{#2}}}
\newcommand{\pdiff}[2]{\frac{\partial{#1}}{\partial{#2}}}
\newcommand{\torate}[1]{\overset{#1}{\to}}
\newcommand{\tofromrate}[2]{\underset{#2}{\overset{#1}{\rightleftharpoons}}}
\newcommand{\tofrom}[0]{\rightleftharpoons}

\begin{document}
\title{Mutual information in time-varying biochemical systems}
\author{Filipe Tostevin} \email{f.tostevin@amolf.nl}
\affiliation{FOM Institute AMOLF, Science Park 104, 1098XG Amsterdam, The
	Netherlands}
\author{Pieter Rein ten Wolde}
\affiliation{FOM Institute AMOLF, Science Park 104, 1098XG Amsterdam, The
	Netherlands}
\date{June 16, 2010}

\begin{abstract}
Cells must continuously sense and respond to time-varying environmental stimuli.
These signals are transmitted and processed by biochemical signalling networks.
However, the biochemical reactions making up these networks are intrinsically
noisy, which limits the reliability of intracellular signalling. Here we use
information theory to characterise the reliability of transmission of
time-varying signals through elementary biochemical reactions in the presence of
noise. We calculate the mutual information for both instantaneous measurements
and trajectories of biochemical systems for a Gaussian model. Our results
indicate that the same network can have radically different characteristics for
the transmission of instantaneous signals and trajectories. For trajectories,
the ability of a network to respond to changes in the input signal is determined
by the timing of reaction events, and is independent of the correlation time of
the output of the network. We also study how reliably signals on different
time-scales can be transmitted by considering the frequency-dependent coherence
and gain-to-noise ratio. We find that a detector that does not consume the
ligand molecule upon detection can more reliably transmit slowly varying
signals, while an absorbing detector can more reliably transmit rapidly varying
signals. Furthermore, we find that while one reaction may more reliably transmit
information than another when considered in isolation, when placed within a
signalling cascade the relative performance of the two reactions can be
reversed. This means that optimising signal transmission at a single level of a
signalling cascade can reduce signalling performance for the cascade as a whole.
\end{abstract}

\maketitle

\section{Introduction}

Cells are continually exposed to a wide range of environmental signals to which
they must react. These stimuli are transmitted and processed within the cell by
networks of proteins and interactions. However, the biochemical reactions which
make up these networks are inherently stochastic events. Recent experiments
have shown that fluctuations associated with this spontaneous reaction noise
can have significant effects on cell phenotype \cite{Elowitz02, Ozbudak02,
Raser04}. In signalling networks, the effect of this inevitable biochemical
noise will be to disrupt the transmission of signals. Random fluctuations mean
that a single input signal can give rise to a distribution of possible
responses. Conversely, a particular response can be generated from a number of
input signals. This uncertainty compromises the ability of a cell to respond
correctly to its environment. It is therefore important to understand how
reliably signals can be transmitted through signalling networks in the presence
of noise. 

A quantitative framework for analysing the reliability of signal communication
in the presence of noise is provided by information theory \cite{Shannon48}.
The application of information theory to neural and sensory signalling has a
long history \cite{Mackay52, Attneave54, Barlow61}, with particular focus on
the reliable encoding of external stimuli in neuronal spiking patterns (for
example \cite{Laughlin81, deRuyter97}).  However, the use of these techniques
in the analysis of intracellular biochemical signalling and gene-regulatory
networks has, until recently, received less attention. In this context a number
of studies have considered the reliability of transmission of signals through
specific reaction systems in the presence of noise \cite{Abshire01, LevineJ07,
Tkacik08a, Tkacik08b, Mehta09} and subject to constraints such as metabolic
cost \cite{dePolavieja04, Tlusty08}. Additionally the impact of network
topology on the transmission of constant signals in generic small gene
regulation and protein signalling motifs has been investigated \cite{Ziv07,
Tkacik09, Walczak09, Mehta09, Walczak10}.

Here we analyse the signalling characteristics of a number of elementary
biochemical reactions for time-varying signals. Specifically, the fidelity of
information transmission between the input and output signals of a biochemical
network is measured by the mutual information. Most previous analyses of the
mutual information for simple reaction motifs \cite{Ziv07, Tkacik08a,
Tkacik08b, Tkacik09, Walczak09, Mehta09, Walczak10} have considered only the
response of a network to signals which do not change on the time-scale of the
network response. For many systems the assumption that the signal is constant
may not be valid. Cells are often exposed to rapidly varying environments, to
which they should also present a time-varying response. Notably, Levine et al
\cite{LevineJ07} calculated the mutual information of an enzymatic push-pull
network in response to an signal pulse. However, they considered only the
information about a restricted two-state input signal that can be extracted
from the instantaneous output level, and did not take into account the
transmission of other dynamical properties of the input. Biochemical networks
often respond not only to the instantaneous values of time-varying signals but
also to other characteristics of the stimulus.  For example, the chemotaxis
network of the bacterium {\em E.  coli} is sensitive to changes in the level of
chemoattractants in the environment, but adapts to constant signals
\cite{Segall86, Block83}. Another example of sensitivity to environmental
changes is the osmotic shock response in budding yeast, where the cell reacts
only to changes in osmotic pressure \cite{Mettetal08}. Cells can also make use
of temporal properties of signals to encode information. For example, in
calcium signalling it is believed that information is encoded in the frequency
and duration of Ca$^{2+}$ bursts, rather than the concentration at any point in
time \cite{Boulware08}. In rat PC-12 cells, stimulation with epidermal growth
factor leads to a transient response of the MAP-kinase pathway, while neuronal
growth factor leads to a sustained response \cite{Marshall95}. In order to
understand the function of these signalling systems we must consider the
transmission of signals which are a function of time, trajectories, through the
network.

We have previously discussed the mutual information rate between trajectories of
biochemical networks \cite{Tostevin09}, and applied these techniques to small
network motifs. This analysis can readily be extended to networks with an
arbitrary number of components and more complex network motifs, such as feedback
\cite{deRonde09a} and feedforward \cite{deRonde09b} loops. However, in order to
make the calculations analytically tractable, even in the simplest cases, we
must make a number of simplifying assumptions. In short, we assume that the
network of interest can be described by a Gaussian model; that is, the
distribution of input or output signal values at any point in time is Gaussian,
as is the conditional distribution between any two points. For these approximate
model systems the mutual information rate can be calculated analytically. For
systems that do not satisfy these assumptions the Gaussian model provides a
lower bound on the rate of information transmission.

In this paper we calculate the mutual information between {\em instantaneous}
signals and the information rate for {\em trajectories} within the Gaussian
approximation for a number of elementary biochemical reactions. We discuss in
detail the assumptions and validity of this model. Our results show that, for
signal trajectories, information is encoded in the timing of reaction events in
signalling cascades. We also compare the performance of different reactions in
isolation and within a simple linear signalling cascade. We find that while a
production reaction can transmit some signals more reliably than an
irreversible conversion reaction, when placed within a signalling cascade
driven by an external upstream signal the relative performance of the two
reactions is reversed. Importantly, this shows that increasing the reliability
of signal propagation for a single step in a cascade does not necessarily
improve, and can in fact degrade, signalling performance for the cascade as a
whole. Our results also show that a detection reaction which irreversibly
consumes its signal substrate can allow for more reliable information
transmission of an upstream signal by reducing noise propagation through the
signalling network.

\section{Formulation: The Gaussian Model} \label{sec:model}

We aim to quantify the performance of biochemical signalling networks by
considering how accurately the network input $S$ can be translated into the
network output $X$. We additionally allow $S$ and $X$ to vary over time, so the
``signal'' which the network is required to transmit is the trajectory $S(t)$
over some time interval, and the output is similarly the trajectory $X(t)$. As a
measure of signalling performance we use the mutual information \cite{Shannon48}
between the input and output trajectories. Formally, the mutual information is
defined in terms of probability distributions over the possible trajectories, 
\begin{multline}
	I(S,X)=\int D[S(t)]\int D[X(t)]\\ p(S(t),X(t))
		\log\left[\frac{p(S(t),X(t))}{p(S(t))p(X(t))}\right]. \label{eq:I_def}
\end{multline}
However a direct evaluation of Eq. \ref{eq:I_def}, either analytically or
numerically, is generally not possible because the space of possible
trajectories is infinite-dimensional. In order to proceed we approximate the
dynamics of the system by a multivariate Gaussian model, for which the mutual
information can be calculated exactly. In this section we discuss the
application of the standard Gaussian communication channel model
\cite{Shannon48} to time-varying biochemical networks, and the assumptions and
approximations which go into this model.

We take as the input and output signals the deviations of $S$ and $X$ from their
average values, $s(t)=S(t)-\langle S(t)\rangle$ and $x(t)=X(t)-\langle
X(t)\rangle$, where $\langle\rangle$ represents averaging over different
realisations of the dynamical system. We assume that $s(t)$ and $x(t)$ are
jointly-Gaussian processes; that is, the joint probability distribution of any
two values of these processes, $p(\alpha(t),\beta(t'))$ for $\alpha,\beta=s$ or
$x$, is a bi-variate Gaussian.

For example, we can construct a vector containing the signal values at discrete
sample times, ${\bf s} =(s(t_1),s(t_2),\dots,s(t_N))$ and similarly for ${\bf
x}$. In general, the trajectories $\mathbf{s}$ and $\mathbf{x}$ can be of
different lengths, $N_s$ and $N_x$. In the Gaussian approximation the joint
distribution of $(\mathbf{s},\mathbf{x})$ is given by 
\begin{equation}
	p(\mathbf{s},\mathbf{x})=(2\pi)^{-(N_s+N_x)/2}|{\bf Z}|^{-1/2}
		\exp\left[-\frac{1}{2}(\mathbf{s}\ \mathbf{x}){\bf Z}^{-1}
			\left(\begin{array}{c}\mathbf{s}\\\mathbf{x}\end{array}\right)
		\right].
	\label{eq:p_gaussian}
\end{equation}
The $(N_s+N_x)\times(N_s+N_x)$ covariance matrix ${\bf Z}$ has the block form
\begin{equation}
	{\bf Z}=\begin{pmatrix} {\bf C}_{ss} & {\bf C}_{xs}\\
		{\bf C}_{sx} & {\bf C}_{xx}\end{pmatrix}, \label{eq:Z}
\end{equation}
where ${\bf C}_{\alpha\beta}$ is an $N_\beta\times N_\alpha$ matrix with
elements given by the correlation function ${\bf C}_{\alpha\beta}^{ij}
=C_{\alpha\beta}(t_j,t_i)=\langle\alpha(t_j)\beta(t_i)\rangle$. Furthermore, the
form of Eq. \ref{eq:p_gaussian} means that the distribution
$p(\alpha(t)|\beta(t'))$ of any value conditional on any other is also Gaussian
with variance $C_{\alpha\beta}(t,t')$.

Shannon \cite{Shannon48} showed that the entropy of an $N-$variate Gaussian
distribution with covariance matrix $\mathbf{C}$ is 
\begin{equation}
	H_G=\frac{1}{2}\log\left[(2\pi e)^N|\mathbf{C}|\right].
\end{equation}
From the definition of the mutual information, therefore, 
\begin{equation}
	I(S,X)=H(S)+H(X)-H(S,X)
	=\frac{1}{2}
		\log\left[\frac{|\mathbf{C}_{ss}||\mathbf{C}_{xx}|}{|\mathbf{Z}|}\right].
	\label{eq:M_det}
\end{equation}
The problem of calculating the mutual information between trajectories in the
Gaussian model reduces to calculating the determinants of the covariance
matrices, a great simplification over the functional integration over ensembles
of trajectories in Eq. \ref{eq:I_def}. In this manuscript we shall consider
two special cases for which the mutual information can readily be evaluated
analytically. Specifically, we shall consider a single instantaneous measurement
of the system, corresponding to $N_s=N_x=1$, and infinite continuous
trajectories, $N_s=N_x\to\infty$ and $t_i-t_{i-1}\to0$. In this latter case the
mutual information rate can be expressed in terms of the spectra of eigenvalues
of the covariance matrices.

One of the crucial assumptions of the Gaussian model is that the input signal is
Gaussian distributed. The validity of this assumption for real biological
systems is not clear, since typical stimulus distributions have not been
measured in most systems. As noted above, we also assume that the marginal or
transfer probabilities between any two signal points is Gaussian. In general,
this is not exactly true for biochemical systems. For systems that do not have
Gaussian statistics, Mitra and Stark \cite{Mitra01} showed that an appropriate
Gaussian model provides a lower bound on the information rate or channel
capacity of the non-Gaussian network. For completeness we reproduce here the
arguments of the proof by Mitra and Stark \cite{Mitra01}:
\begin{enumerate}
	\item The channel capacity subject to a power constraint on $S$ is defined as
		$C(S,X)=\max_{p(s)}I(S,X)$, where the maximisation is over all input
		distributions satisfying the constraint. 
	\item We can construct a Gaussian input distribution on $S$ satisfying the
		power constraint, $p_G(s)$. From the definition of the channel capacity,
		$C(S,X)\geq I(S_G,X)$, since the Gaussian is not necessarily the optimal
		input distribution for the channel.
	\item We can also construct a multivariate Gaussian model with the same second
		moments as the non-Gaussian system when the input distribution is chosen to
		be $p_G(s)$. The mutual information in this case $I(S_G,X_G)\leq I(S_G,X)$.
		This is essentially because a Gaussian distribution has the largest entropy
		for a given variance, such that a Gaussian transfer function maximises the
		uncertainty of the output for a given input.
 \item In summary, $I(S_G,X_G)\leq I(S_G,X)\leq C(S,X)$.
\end{enumerate}
For a Gaussian system, the mutual information can be calculated exactly, and the
mutual information equals the channel capacity. For systems with a Gaussian
input distribution but which are otherwise non-Gaussian, the mutual information
is bounded from below by the mutual information for the Gaussian model {\em with
the same second moments as the non-Gaussian system}. For a general non-Gaussian
system with a power constraint, the mutual information calculated in this way is
a lower bound on the channel capacity.

Van Kampen's linear noise approximation (LNA) \cite{VanKampen} provides a
prescription by which we can approximate a network of interest by one which
satisfies the requirements of the Gaussian model. In this approximation we
assume that the intrinsic noise in the network is Gaussian-distributed and small
relative to the mean, and we linearise the network response around steady state.
For linear systems it is known that the second moments which are calculated in
the LNA are exact \cite{Warren06}. Thus for a linear system the LNA can be
used to estimate a lower bound on information transmission, which becomes exact
in the limit of small Gaussian noise. However, for non-linear systems we are not
guaranteed that the second moments calculated in the LNA are the same as those
of the full non-linear system. Therefore, the LNA does not necessarily lead to
an appropriate Gaussian model of the network in the sense of providing a bound
on the information rate. For some non-linear systems the LNA has also been found
to provide an accurate description of the second moments of networks
\cite{Ziv07, TanaseNicola06, Bruggeman09, deRonde09a}, particularly for systems
with large molecular copy numbers where non-linear effects are negligible.
Crucially, though, it is not necessarily obvious {\em a priori} for a given
non-linear network whether or not the LNA will provide an accurate model. Hence
if one wishes to consider non-linear networks in the LNA, it is important to
verify that the second moments of the approximate model system match those of
the full network, for example with stochastic simulations. In the remainder of
this paper we will focus only on linear systems for which the second moments can
be calculated exactly.

In the model as described above there is no requirement that the system is in a
macroscopic steady state  
\footnote{For the purposes of this discussion, a system is considered to be in
	steady state if it satisfies $\langle S(t)\rangle=\langle S(0)\rangle$ and
	$\langle S(t)S(t')\rangle=\langle S(t-t')S(0)\rangle$ for all $t$ and $t'$,
	and similarly for $X$.}.
However, henceforth we shall assume that our systems are in such a steady-state
as this simplifies the calculation of the correlation functions. Additionally in
steady state the correlation functions depend only on time-differences,
$C(t,t')=C(t'-t)$, which restricts the form of the covariance matrices and
facilitates calculation of the determinants. Furthermore, the assumption of
steady state simplifies the interpretation of the calculated information values.

\subsection{Mutual information between instantaneous measurements}

In this section we discuss the mutual information between the instantaneous
value of the output signal, $X(t_0)$, and the input signal at the same point in
time, $S(t_0)$. The instantaneous mutual information tells us, if we know the
output of the network at a particular time, how much we learn about the current
state of the input process. We note that this differs from previous analyses of
the mutual information for {\em constant} signals \cite{Ziv07, Tkacik08a,
Tkacik08b, Tkacik09, Walczak09, Mehta09, Walczak10}, because $S$ remains a
dynamic variable which changes over time. In particular, we take into account
the fact that the correlation timescale of the input signal and the response
time of the processing network are both finite. The interplay of changes in the
signal and response will be particularly important when these timescales are
comparable.  The instantaneous mutual information considers not only the
``intrinsic'' noise in $X$ at constant $S$ due to the stochastic nature of the
production and decay reactions, but also the ``extrinsic'' variability in $X$
which arises from changes in the input signal.

An instantaneous measurement of the input and output signals represents a
special case of the Gaussian model in which the input and output vectors are
each one-dimensional,
\begin{eqnarray}
	p(s,x)=\frac{1}{2\pi|\mathbf{Z}|^{1/2}}\exp\left[-\frac{1}{2}\left(s\ x\right)
		\mathbf{Z}^{-1}\left(\begin{array}{c} s \\ x \end{array}\right)\right],
	\label{eq:gauss_inst}
\end{eqnarray}
and the elements of the matrix $\mathbf{Z}$ are the instantaneous covariances,
$\sigma_{\alpha\beta}^2=\langle\alpha(t_0)\beta(t_0)\rangle$:
\begin{equation}
	\mathbf{Z}=\left(\begin{array}{cc} \sigma_{ss}^2 & \sigma_{sx}^2 \\
		\sigma_{sx}^2 & \sigma_{xx}^2 \end{array}\right).
\end{equation}
Since we allow both the input and output signals to vary in time, these
covariances include the interplay of the input and output signal timescales.
From Eq. \ref{eq:gauss_inst} it follows that the conditional distribution of $x$
given $s$ is 
\begin{eqnarray}
	p(x|s)=\frac{1}{(2\pi\sigma_{x|s}^2)^{1/2}}\exp\left[-\frac{(x-\langle
		x|s\rangle)^2}{2\sigma_{x|s}^2}\right],
\end{eqnarray}
where $\langle x|s\rangle={\sigma_{sx}^2}s/{\sigma_{ss}^2}$ and
$\sigma_{x|s}^2={|\mathbf{Z}|}/{\sigma_{ss}^2}$. We can therefore define the
network gain $g=\sigma_{sx}^2/\sigma_{ss}^2$ and intrinsic noise
$\sigma_{x|s}^2=\sigma_{xx}^2-g^2\sigma_{ss}^2$, and the output signal of the
network takes the form $x=gs+\eta_{x|s}$, where $\eta_{x|s}$ is a
Gaussian-distributed random variable with variance $\sigma_{x|s}^2$. This is
the canonical form of the additive white Gaussian noise channel considered by
Shannon and others \cite{Shannon48, Fano, Cover}. We note that in general the
gain $g$ defined above differs from the ``macroscopic'' gain in the steady
state values of $S$ and $X$, $\pdiff{\langle X\rangle}{\langle S\rangle}$,
which characterises the transmission of constant signals.

From the definition of the mutual information, 
\begin{eqnarray}
	I_{\rm inst}(S,X)&=&\frac{1}{2}\log\left(\frac{\sigma_{ss}^2\sigma_{xx}^2}
		{|\mathbf{Z}|}\right)
	=\frac{1}{2}\log\left(1+\frac{\sigma_{sx}^4}{|\mathbf{Z}|}\right)
	\label{eq:m_inst_shannon}
\end{eqnarray}
By comparison of Eq. \ref{eq:m_inst_shannon} with the well-known result for the
capacity of a Gaussian channel \cite{Shannon48}, we can interpret this result
in terms of signal-to-noise ratio:
\begin{equation}
	\frac{{\rm Signal}}{{\rm Noise}}=\frac{\sigma_{sx}^4}{|\mathbf{Z}|}
		=\frac{g^2}{\sigma_{x|s}^2}\sigma_{ss}^2.
\end{equation}
In essence, to reliably detect an input signal $s$ the network gain has to raise
the output signal $x$ above the noise level $\sigma^2_{x|s}$. The gain-to-noise
ratio $g^2/\sigma_{x|s}^2$ provides a signal-independent measure of the
performance of the network. For a given input signal, we therefore expect that
the mutual information is maximised when the ratio $g^2/\sigma^2_{x|s}$ is
maximised. The gain-to-noise ratio is also the Fisher information \cite{Cover}
about the signal $s$ contained in the sample $x$. For a Gaussian system this is
the reciprocal of the average error in estimating $s$ from a given output $x$. 

\subsection{Mutual information rate for infinite trajectories}
\label{sec:trajectories}

A second special case of the Gaussian model is the limit of infinite,
continuous, trajectories. We define the entropy rate of a Gaussian process as 
\begin{equation}
	h_G=\lim_{N\rightarrow\infty}\frac{H_G}{N\Delta}, \label{eq:rate}
\end{equation}
where $\Delta$ is the sampling interval of the signal and $T=N\Delta$ is the
length of the trajectory. Since we assume that the network is fluctuating
around steady state, the covariance between the input (or output) signal at two
time points depends only on the time difference between the two samples. The
corresponding covariance matrix ($\mathbf{C}_{ss}$ or $\mathbf{C}_{xx}$)
therefore has a Toeplitz structure, which allows us to rewrite the matrix
determinant in the limit $N\rightarrow\infty$ in terms of the power spectrum of
the signal $P(\omega)$ \cite{Fano, Cover},
\begin{equation}
	h_G=\frac{1}{2\Delta}\log(2\pi e)
		+\frac{1}{4\pi}\int_{-\omega_0}^{\omega_0}d\omega\log P(\omega),
		\label{eq:entropyrate_spectrum}
\end{equation}
where $\omega_0=\pi/\Delta$ is the angular Nyquist frequency. In effect we have
decomposed the signal into an infinite number of independent frequency
components with corresponding variance $P(\omega)$. Note that in the limit
$\Delta\rightarrow0$ the entropy rate is not well defined. However, as we will
see below, the mutual information rate can remain finite in this limit.

To calculate the mutual information we also need the joint entropy rate
$h(S,X)$. The covariance matrix ${\bf Z}$ for the combined $(\mathbf{s},
\mathbf{x})$ signal is not (in general) Toeplitz. However, following
\cite{Munakata}, the joint entropy rate for a Gaussian system can be written in
terms of the (cross-) power spectra of the input and output signals,
\begin{multline}
	h(S,X)=\frac{1}{\Delta}\log(2\pi e) 
		+\frac{1}{4\pi}\int_{-\omega_0}^{\omega_0}d\omega \\
		\log[P_{ss}(\omega)P_{xx}(\omega)-|P_{sx}(\omega)|^2],
\end{multline}
where $P_{\alpha\beta}(\omega)$ is the Fourier transform of
$C_{\alpha\beta}(t)$. Combining this expression with Eq.
\ref{eq:entropyrate_spectrum} and taking the limit $\Delta\rightarrow0$, the
mutual information rate between continuous trajectories can be written as
\begin{eqnarray}
	R(S,X)&=&h(S)+h(X)-h(S,X) \label{eq:R_def} \\
		&=&-\frac{1}{4\pi}\int_{-\infty}^{\infty} d\omega
		\log\left[1-\frac{|P_{sx}(\omega)|^2}{P_{ss}(\omega)P_{xx}(\omega)}\right]
	\label{eq:I_S_omega}.
\end{eqnarray}

We can recognise that the total information rate is the sum of independent
Gaussian channels at different frequencies. The output trajectory of the network
is, by construction, a stationary Gaussian process. For such a process it is
known that the different Fourier components must be statistically independent
\cite{Fano}. In order for this to be realised in a signalling system, we
require that an input signal at a specific frequency leads to an output only at
a unique frequency; if the network has a response at multiple frequencies, the
components of the output at these frequencies will be correlated. This is
equivalent to assuming that the network response is linear. Furthermore, the
noise in the network must be purely additive. As described above, the LNA
provides a prescription for constructing a model system which satisfies these
requirements. For systems that are non-linear or non-Gaussian, Eq.
\ref{eq:I_S_omega} provides a bound on the information rate or channel capacity,
provided that the power spectra of the full non-linear system are used in the
calculation of $R(S,X)$.

How should we interpret the ``rate'' $R(S,X)$? From Eqs. \ref{eq:rate} and
\ref{eq:R_def} we see that $R(S,X)=\lim_{T\to\infty} I(S,X;T)/T$, the total
mutual information for two long trajectories divided by the trajectory length.
Thus, $R(S,X)$ is the average mutual information per unit time for a long
trajectory. From this definition it also follows that
$R(S,X)=\lim_{T\rightarrow\infty}\diff{I(S,X;T)}{T}$; $R(S,X)$ is the rate at
which we gain information about the input trajectory as the length of both the
input and output trajectories is increased. However, some care should be taken
with this interpretation. If we record an output trajectory of length $T$, and
then measure for an additional time $T'$, the total information we have about
the input trajectory as a whole will increase by approximately $T'R(S,X)$.
However, the additional information we have gained will not be restricted to the
new segment of the input signal, but will also be distributed over the original
trajectory. Therefore, it is not correct to say that we have learnt $T'R(S,X)$
about the segment $T\leq t\leq T+T'$ of the input trajectory. More generally,
there may be additional contributions to the mutual information which are
important for short trajectories. When $T$ is comparable to the correlation
times of the network we would expect $I(S,X;T)/T$ to deviate from $R(S,X)$.

The ratio $\phi(\omega)=\frac{|P_{sx}(\omega)|^2}{P_{ss}(\omega)P_{xx}(\omega)}$
which appears in Eq. \ref{eq:I_S_omega} is the {\em coherence} between $s(t)$
and $x(t)$. This is a standard measure of the correlation between the signals
$s$ and $x$, with $\phi(\omega)=0$ for independent signals and $\phi(\omega)=1$
when $s$ completely determines $x$. We can also define the signal and noise
power spectra via analogous expressions to those for instantaneous measurements,
\begin{eqnarray}
	\Sigma(\omega)&=&g^2(\omega)P_{ss}(\omega)=\frac{|P_{sx}(\omega)|^2}
		{P_{ss}(\omega)} \label{eq:power} \label{eq:signal_def} \\
	N(\omega)&=&P_{xx}(\omega)-\Sigma(\omega) \label{eq:noise}.
\end{eqnarray}
We can see that the coherence represents the signal fraction of the total output
power, $\phi(\omega)=\Sigma(\omega)/[\Sigma(\omega)+N(\omega)]$. The mutual
information rate can also be written in terms of the signal-to-noise ratio,
\begin{equation}
	R(S,X)=\frac{1}{4\pi}\int_{-\infty}^{\infty} d\omega
		\log\left[1+\frac{\Sigma(\omega)}{N(\omega)}\right] \label{eq:I_sn_omega},
\end{equation}
recovering the usual expression for the capacity of a continuous Gaussian
channel \cite{Fano,Cover}.

From \cite{Tostevin09} and above we can see that the ability of a network to
transmit information in a time-varying signal at a particular frequency is
characterised by the frequency-dependent gain-to-noise ratio,
$g^2(\omega)/N(\omega)$. To understand signalling performance it is therefore
important to consider both the gain and the noise of the network, and not simply
the gain or noise in isolation. Furthermore, for systems which satisfy the
spectral addition rule \cite{TanaseNicola06}, meaning that the network dynamics
do not affect the input signal itself, the gain-to-noise ratio is
signal-independent and characterises the intrinsic transmission characteristics
of the network. Knowledge of the gain-to-noise ratio also allows us to calculate
the optimal signal that maximises the information rate for a given network,
through the ``water-filling'' approach of Fano \cite{Fano}. Finally, in the same
way as for instantaneous measurements, the gain-to-noise ratio at a specific
frequency is the Fisher information provided by the output signal $x(\omega)$
about the input signal $s(\omega)$ at this frequency. Thus, the gain-to-noise
ratio is the reciprocal of our uncertainty in estimating the input signal given
a particular output.

We saw previously that the entropy rate of a stochastic process in the continuum
limit is not well defined, but the mutual information rate can still be
calculated. Under what conditions do we find a finite mutual information rate?
From Eqs \ref{eq:I_S_omega} and \ref{eq:I_sn_omega} we can see that the mutual
information rate will be divergent if the signal-to-noise ratio
$\Sigma(\omega)/N(\omega)$ or coherence $\phi(\omega)$ do not approach zero as
$\omega\to\infty$. The power spectra of biochemical reactions typically take the
form of rational polynomials in $\omega^2$ (as we shall see in the following
sections). Then the integral in the mutual information rate will converge to a
finite value if the signal power decreases more rapidly at high frequencies than
the noise power. This characteristic form of the power spectra also often allows
us to perform the integral in Eq. \ref{eq:I_S_omega} or \ref{eq:I_sn_omega} and
calculate an explicit expression for $R(S,X)$. To do this we make use of the
result that 
\begin{equation}
	\int_0^\infty d\omega
		\ln\left[\frac{\omega^2+a^2}{\omega^2+b^2}\right]=\pi(a-b).
\end{equation}
Using the properties of the logarithm, this can trivially be extended to an
arbitrary number of terms, 
\begin{equation}
	\int_{-\infty}^\infty d\omega \ln\left[\prod_{i=1}^N\frac{\omega^2+a_i^2} 
		{\omega^2+b_i^2}\right]
	=2\pi\left[\sum_{i=1}^Na_i-\sum_{i=1}^Nb_i\right].
	\label{eq:w2_int}
\end{equation}
We emphasise that for the integral to be cast in this form we require that the
coefficient of the leading order term in both the denominator and numerator of
Eq. \ref{eq:w2_int} is 1. Then calculating $R(S,X)$ reduces to finding the roots
of polynomials which are constructed from the network power spectra.

Can we still compare the performance of networks with a divergent information
rate? In these networks the coherence and signal-to-noise ratio approach to some
non-zero value as $\omega\rightarrow\pm\infty$. Rather than Eq. \ref{eq:w2_int},
the integral in the expression for the mutual information rate will have the
form 
\begin{multline}
	\int_{-\omega_0}^{\omega_0} d\omega \ln\left[k\prod_{i=1}^N
		\frac{\omega^2+a_i^2}{\omega^2+b_i^2}\right]
	=2\omega_0\ln k+\\
		2\omega_0\ln\left[\prod_{i=1}^N\frac{\omega_0^2+a_i^2}
			{\omega_0^2+b_i^2}\right]\\ +
		4\sum_{i=1}^N\left(a_i\arctan\frac{\omega_0}{a_i}-
			b_i\arctan\frac{\omega_0}{b_i}\right).
\end{multline}
In the limit of $\Delta\rightarrow0$ and hence
$\omega_0=\pi/\Delta\rightarrow\infty$ the last two terms above can be
neglected. Information transmission for these networks is therefore dominated by
high-frequency components, and can be characterised by the constant $k$. The
values of $k$ for different networks can therefore be used to compare their
relative information rates.

\section{Results} \label{sec:results}

In this section we present results for some elementary molecular reactions, 
considered previously \cite{Tostevin09, TanaseNicola06} and summarised in Table
\ref{tab:motifs}. These reaction schemes are significant because they exemplify
the three basic ways in which $S$ can directly drive the production of $X$:
reversible conversion between $S$ and $X$; irreversible conversion from $S$ to
$X$; and stimulating production of $X$ without consuming an $S$ molecule. Since
these schemes feature only first-order reactions the covariances and power 
spectra can be calculated exactly from the chemical master equation 
\cite{Warren06}.

\begin{table}
\begin{tabular}{cc}
Motif & Reactions
\\[3pt]\hline\\[-10pt]
	(I)
& $\begin{array}{c}
		\begin{array}{cc} 
			\emptyset \xrightarrow{\kappa} S & S\xrightarrow{\lambda} \emptyset
		\end{array} \\ 
		S\tofromrate{\rho=kW}{\mu} X 
	\end{array}$
\\[5pt]\hline\\[-10pt]
	(II) 
& $\begin{array}{cc} 
		\emptyset \xrightarrow{\kappa} S & S\xrightarrow{\lambda} \emptyset \\ 
		S\xrightarrow{\rho} X & X\xrightarrow{\mu} \emptyset
	\end{array}$
\\[5pt]\hline\\[-10pt]
	(III)
& $\begin{array}{cc} 
		\emptyset \xrightarrow{\kappa} S & S\xrightarrow{\lambda} \emptyset \\ 
		S\xrightarrow{\rho} S+X & X\xrightarrow{\mu} \emptyset
	\end{array}$
\\[5pt]\hline
\end{tabular}
	\caption{Summary of the three elementary reaction motifs. In motif I we assume
		that the number of $W$ proteins is sufficiently large that the rate $\rho$
		can be taken to be constant.}
 \label{tab:motifs}
\end{table}

\subsection{Instantaneous mutual information}

\subsubsection{Motif I: Reversible binding}

This motif describes the association and disassociation of a signalling
molecule, $S$, and a receptor, $W$, to form an active complex, $X$. We assume
that $W$ is present at high copy numbers, such that its depletion due to binding
$S$ can be neglected. In this case we consider the ``signal'' we wish to detect
to be the total number of both bound and unbound signalling molecules,
$S_T(t_0)=S(t_0)+X(t_0)$. For this motif the covariances are
\begin{equation}
	\begin{array}{ccc}
		\sigma_{s_Ts_T}^2=\langle S_T\rangle, & 
		\sigma_{s_Tx}^2=\sigma_{xx}^2=\langle X\rangle,
	\end{array}
\end{equation}
which gives for the instantaneous mutual information
\begin{equation}
	I_{\rm inst}(S_T,X)=-\frac{1}{2}\log\left(1-\frac{\rho}{\rho+\mu}\right)
		=-\frac{1}{2}\log\left(1-\frac{\langle X\rangle}{\langle S_T\rangle}\right).
\end{equation}
The instantaneous mutual information between $X$ and $S_T$ is determined simply
by the fraction of $S_T$ which is in the bound ($X$) state on average. Indeed
the average statistics of this binding reaction are simply those of a binomial
distribution. Each molecule in the system can be in two states: the bound $X$
state with probability $\rho/(\rho+\mu)=\langle X\rangle/\langle S_T\rangle$, or
the unbound $S$ state with probability $\langle S\rangle/\langle S_T\rangle$.
Since each molecule is independent, if there are $N_{S_T}$ molecules in the
system in total, the expected number of $X$ molecules will be $N_{S_T}\langle
X\rangle/\langle S_T\rangle=gN_{S_T}$, and the variance in the number of $X$
molecules will be $N_{S_T}g(1-g)$. Averaging over all possible values of $S_T$,
the intrinsic noise in $X$ is $\sigma_{x|s_T}^2=\langle S_T\rangle g(1-g)$. For
instantaneous signals, the time-scales of the binding and dissociation reactions
are not important for information transmission, but only their ratio. This is
because we wish to estimate only the current state of the system, and we are not
concerned with how rapidly the state of the system changes.

\subsubsection{Motif II: Irreversible modification}

This motif is characterised by the irreversible conversion of an $S$ molecule to
an $X$, $S\to X$. Such a reaction could represent irreversible
post-transcriptional modification of a protein such as cleavage, or binding of a
ligand to a receptor followed by rapid endocytosis of the resulting complex. For
this motif the covariance $\sigma_{sx}^2=0$. The number of $X$ molecules present
in the system depends only on how many $S$ molecules have been converted to $X$
in the past. If the production of different $S$ molecules occurs independently,
then the instantaneous values of $S$ and $X$ are uncorrelated since the $S$
molecules that have previously decayed are uncorrelated from those that are
currently present in the system. Hence the instantaneous mutual information
between a single measurement of $X$ and the simultaneous value of $S$ is zero;
measuring $X$ tells us nothing about about how many $S$ molecules are currently
present in the system.

\begin{figure}
\includegraphics{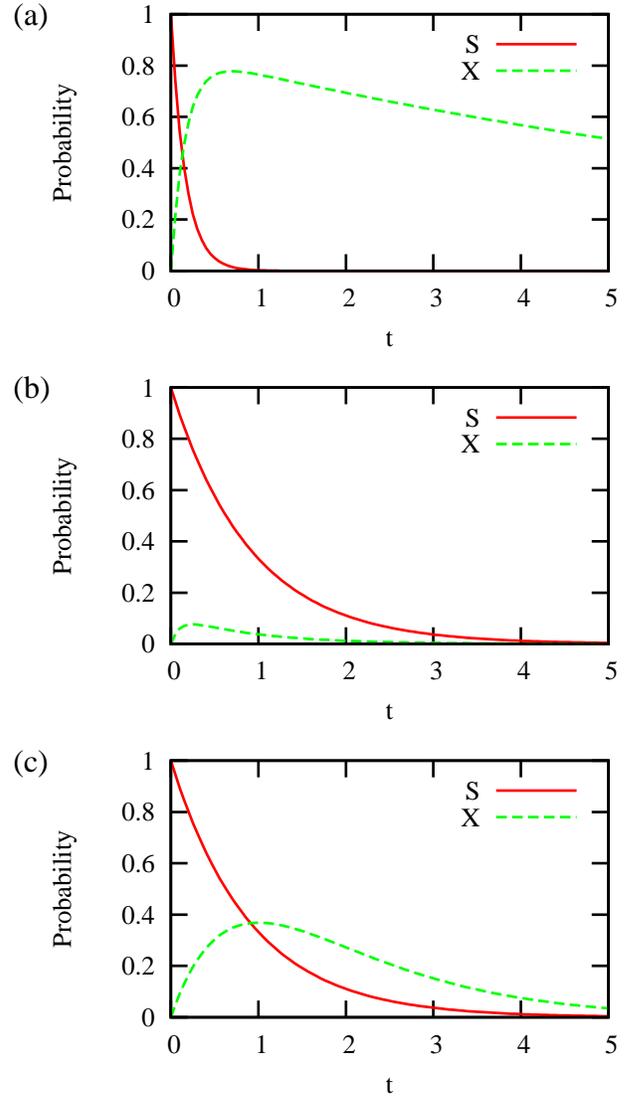}
\caption{Probability of a molecule produced at $t=0$ being in
	each molecular state of motif II as a function of time for different
	reaction rates.  (a) When $\rho\gg\mu,\lambda$, $S$ molecules rapidly
	decay to $X$, but for most of the lifetime of $X$ the chance of an $S$
	molecule from the same burst being present is small. Parameters:
	$\mu=0.1$, $\lambda=1$, $\rho=5$.  (b) When $\mu\gg\rho,\lambda$, $S$
	molecules which decay to $X$ are rapidly degraded. Therefore the
	probability of observing an $X$ molecule is small, and little
	information can be gained. Parameters: $\mu=10$, $\lambda=0.1$,
	$\rho=1$.  (c) Information is maximised when there is a significant
	probability of finding both $S$ and $X$ molecules. The figure shows an
	optimal choice of $\rho$ for burst size $n=2$, $\mu=0.9$,
	$\lambda=0.1$: $\rho=1$.}
\label{fig:motif2burst}
\end{figure}

Note however that if there are correlations between the production of different
$S$ molecules, such as if $S$ molecules are produced in bursts, then the
instantaneous $S$ and $X$ values become correlated and the instantaneous mutual
information will be non-zero. Suppose that $S$ molecules are produced instead
via the reaction $\emptyset\xrightarrow{\kappa}nS$. Then the mutual information
between instantaneous values of $S$ and $X$ is given by 
\begin{multline}
	I_{\rm inst}(S,X)=-\frac{1}{2}\\ 
	\log\left(1-\frac{(n-1)^2\rho\mu}
		{(n+1)(\lambda+\mu+\rho)(2\lambda+2\mu+(n+1)\rho)}\right).
\end{multline}
Interestingly, for a fixed production rate of $X$, $\rho$, the mutual
information
can be optimised by choosing 
\begin{equation} \label{eq:inst2_mu_opt}
	\mu_{\rm opt}=\sqrt{(\lambda+\rho)(\lambda+\frac{n+1}{2}\rho)},
\end{equation}
while for a fixed degradation rate of $X$, $\mu$, information is maximised for
\begin{equation} \label{eq:inst2_rho_opt}
	\rho_{\rm opt}=(\lambda+\mu)\sqrt{\frac{2}{n+1}}.
\end{equation}
These optima are the result of a trade-off between the probability that when we
observe an $X$ molecule other $S$ molecules produced in the same burst are still
in the system, and the probability of observing an $X$ molecule at all, as
depicted in Fig. \ref{fig:motif2burst}. For example, if the production rate of
$X$ is too large, all $S$ molecules rapidly decay to $X$. Since the probability
of $S$ molecules from the same burst remaining in the system is low, we lose the
ability to predict the value of $S$. If $\rho$ is too small, the chance of a
single $X$ molecule being produced in a burst becomes small and we again lose
information, this time about the bursts which go undetected. Similar arguments
apply for $\mu$: if $\mu$ is too small then over the lifetime of $X$ the $S$
molecules from the same burst will typically either have been degraded or also
have decayed to $X$; if we make $\mu$ too large then we very rarely find an $X$
in the system, and hence on average we gain little information about $S$. Note
that Eqs \ref{eq:inst2_mu_opt} and \ref{eq:inst2_rho_opt} cannot both be
satisfied simultaneously - there are no isolated optimal parameter combinations.

\subsubsection{Motif III: Production}

This motif may represent an effective coarse-grained model of an enzymatic
reaction or of protein production in which fast reaction steps have been
integrated out. For this motif the covariances are:
\begin{eqnarray} 
	\begin{array}{ccc}
		\sigma_{ss}^2=\langle S\rangle, &
		\sigma_{sx}^2=\frac{\rho\langle S\rangle}{\lambda+\mu}, &
		\sigma_{xx}^2=\langle X\rangle\left(1+\frac{\rho}{\lambda+\mu}\right).
	\end{array}
\end{eqnarray}
The gain is given by $g=\rho/(\lambda+\mu)$, and describes the average response
of the output signal $X$ to a perturbation in the input signal. Suppose we
initially have no $S$ or $X$ molecules present in the system. We then introduce
a single $S$ molecule at $t=0$. The survival probability for this $S$ molecule,
the probability that it has not yet decayed at time $t$, will be exponentially
distributed with a mean lifetime of $1/\lambda$. If the $S$ molecule has not
decayed at $t$, the mean number of $X$ molecules present will be $\bar{X}(t)
=\rho[1-\exp(-\mu t)]/\mu$. Now we observe the system at a time $t>0$
chosen according to the probability that the $S$ molecule is still present. This
time will then be drawn from the distribution $p(t)=\exp(-\lambda
t)/\left[\int_0^{\infty}\exp(-\lambda t)dt\right] =\lambda\exp(-\lambda t)$, and
the mean number of $X$ molecules that we will observe is $\int_0^{\infty}
p(t)\bar{X}(t)dt=\rho/(\lambda+\mu)$. We can therefore see that the gain
directly measures the typical change we observe in the number of $X$ molecules
if the input signal changes by one $S$ molecule. 

The instantaneous mutual information is given by
\begin{equation}
	I_{\rm inst}(S,X)=-\frac{1}{2}\log\left[1-\frac{\rho\mu}{
		(\lambda+\mu)(\lambda+\mu+\rho)}\right].
\end{equation}
We see that $I_{\rm inst}(S,X)$ has a maximum as a function of $\mu$ at 
\begin{equation}
	\mu_{\rm opt}=\lambda\sqrt{1+\frac{\rho}{\lambda}}.
\end{equation}
This optimal value appears as a result of the interplay between intrinsic noise
and temporal correlations in the system. When $\mu\gg\lambda$, intrinsic
fluctuations in $X$ are much more rapid than the systematic changes in $X$ due
to variations in $S$. In this case, the instantaneous statistics of $X$ are
approximately those of a simple Poisson birth-death process in response to a
constant $S$ input. The gain is $g=\rho/(\mu+\lambda)\approx\rho/\mu=\langle
X\rangle/\langle S\rangle$, the number of $X$ molecules per $S$ which would be
observed for a constant $S$ signal. The noise variance in $X$ is approximately
that of a Poisson process, $\sigma_{x|s}^2\approx\langle X\rangle\sim\mu^{-1}$.
While increasing $\mu$ decreases the absolute noise strength, the relative noise
in the mean $\sigma_{x|s}^2/\langle X\rangle^2$ increases. Since signalling
fidelity depends on the ratio $g^2/\sigma_{x|s}^2\propto\langle X\rangle^2
/\sigma_{x|s}^2$, increasing $\mu$ decreases the gain-to-noise ratio and reduces
the transmitted information. In the opposite limit, when $\mu\ll\lambda$, the
output signal $X$ effectively integrates over variations in $S$. In this case
the gain is $g=\rho/(\mu+\lambda)\approx\rho/\lambda$, the mean number of $X$
molecules produced during the lifetime of an $S$ molecule. Since the typical
lifetime of $X$ is long compared to that of $S$ we can assume that no $X$
molecule decays before the $S$ from which it was produced. Once we are in this
regime further decreasing $\mu$ has no effect on our ability to amplify the
incoming signal, which is instead limited by the lifetime of $S$; integrating
for a longer time provides no further benefit. The intrinsic noise variance,
however, is still proportional to $\langle X\rangle$ and hence increases as
$\mu$ is decreased. As a result, for small $\mu$ the gain-to-noise ratio goes as
$g^2/\sigma_{x|s}^2\sim\mu$. The precise value of the optimal decay rate also
depends on the production rate of $X$, $\rho$, since this determines the
time-scale of fluctuations in $S$ to which $X$ can respond.

\subsection{Mutual information rate}

For the reactions shown in Table \ref{tab:motifs} the power spectra can readily
be calculated \cite{TanaseNicola06, Warren06}. Some results for these motifs
were presented in \cite{Tostevin09}. Here we extend the discussion of these
results.

\subsubsection{Motif I}

For reaction motif I we again consider the input signal to be
$s_T(t)=s(t)+x(t)$. For this motif, the gain-to-noise ratio between the signals
$s_T(t)$ and $x(t)$ is given by 
\begin{equation}
	\frac{g^2(\omega)}{N(\omega)}=\frac{\rho}{2\kappa}
		\frac{\lambda(\mu+\lambda+\rho)^2}{[\omega^2+(\mu+\rho)^2+\rho\lambda]}.
\end{equation}
We see that information capacity of the network decreases at high frequencies.
This network motif effectively acts as a low-pass filter for information. The
binding and unbinding reactions cannot track extremely rapid changes in $S$ or
$X$, and therefore high-frequency components of the $s_T(t)$ signal are not
transmitted to $x(t)$.

The mutual information rate for this motif can be calculated by performing the
integral in Eq. \ref{eq:I_S_omega}, and is given (in nats per unit time) by
\begin{equation}
	R(S_T,X)=\frac{\lambda}{2}\left[\sqrt{1+\beta}+\sqrt{\beta+[\beta+\alpha]^2}-
		(1+\alpha+\beta)\right],
\end{equation}
where $\alpha=\mu/\lambda$ and $\beta=\rho/\lambda$ are respectively the rates
of the dissociation and association reactions relative to the lifetime of $S$.
The mutual information rate increases with increasing $\beta$ and decreases with
increasing $\alpha$. Therefore, as for instantaneous measurements, the amount of
information about $s_T(t)$ which we can extract from the trace $x(t)$ increases
as the average fraction of molecules in the $X$ state, which is determined by
$\rho/\mu=\beta/\alpha$, increases. However, unlike the instantaneous mutual
information, the information rate between trajectories does depend on the
absolute binding and unbinding rates and not just their ratio. This is because
the mutual information rate takes into account the timescale on which the number
of $X$ molecules is able to track changes in the number of $S$ molecules. It
should also be noted that changing $\rho$ or $\mu$ also affect the statistics of
the input signal $s_T(t)$, since molecules are protected from degradation while
in the $X$ state. This means that the entropy of the input distribution will
change as $\rho$ or $\mu$ are varied. 

\subsubsection{Motif II}

The coherence for this motif is $\phi(\omega)=\rho/4(\rho+\lambda)$, independent
of $\omega$. The integral in Eq. \ref{eq:I_S_omega} is therefore divergent,
giving an infinite information rate. In reality, the integral of the mutual
information rate should be truncated at some large but finite frequency, since
the biochemical reactions which we have modelled as instantaneous jump processes
actually take some finite time to occur. Nevertheless, we can conclude that
observing the trajectory $x(t)$ provides a large amount of information about the
trajectory $s(t)$ because for every production event of $X$ we know when one $S$
molecule has decayed. However, we do not have a complete knowledge of the input
signal $s(t)$ and thus the coherence remains less than 1. We can see that for
$\lambda\gg\rho$, $\phi\rightarrow0$. Since the majority of $S$ molecules are
degraded directly, and do not decay via the $X$ form, most of the molecules
which pass through the system are never observed in the output signal. Hence the
fraction of the input signal which we measure will be very small. Conversely,
when $\rho\gg\lambda$, $\phi\approx1/4$. In this case we observe the decay of
all $S$ molecules, but we still have some uncertainty about their production
times.

We note also that signalling fidelity, as determined by either the coherence or
signal-to-noise ratio, is independent of the decay rate of X, $\mu$. Since decay
events of $X$ occur independently of the number of $S$ molecules present, they
contribute no information about the input signal. We can understand this by
considering the timing of production and decay events in the trajectory of $X$.
Indeed, the trajectory of the number of $X$ molecules as a function of time will
consist of discrete steps at which $X$ molecules are produced and decay. Apart
from a constant offset which can be absorbed into $\langle X\rangle$, the
trajectory $x(t)$ is completely described by the sequence of times at which the
number of $X$ molecules increase and decrease, $\lbrace t_+\rbrace$ and $\lbrace
t_-\rbrace$ respectively. We can therefore write the probability of a particular
trajectory as $p(x(t))=p(\lbrace t_+\rbrace,\lbrace t_-\rbrace)= p(\lbrace
t_+\rbrace)p(\lbrace t_-\rbrace|\lbrace t_+\rbrace)$. The production of $X$
molecules is regulated by the signal $S$, but their decay is not. That is, the
timing of production events is dependent on the input $s(t)$, but the timing of
decay events is determined solely by the intrinsic dynamics of $X$ and therefore
does not depend on $S$ explicitly. In this case, we can also factorise the
conditional probability of a given trajectory $p(x(t)|s(t))=p(\lbrace
t_+\rbrace|s(t)) p(\lbrace t_-\rbrace|\lbrace t_+\rbrace)$. The mutual
information can then be written as
\begin{multline}
	I(s(t),x(t))=\int D[s(t)]\int D\lbrace t_+\rbrace\int D\lbrace t_-\rbrace\\
		p(s(t),\lbrace t_+\rbrace,\lbrace t_-\rbrace) \log\left[\frac{
			p(s(t))p(\lbrace t_+\rbrace|s(t))p(\lbrace t_-\rbrace|\lbrace t_+\rbrace)}
			{p(s(t))p(\lbrace t_+\rbrace)p(\lbrace t_-\rbrace|\lbrace t_+\rbrace)}
		\right],
\end{multline}
where $D\lbrace t\rbrace$ represent integration over all possible sequences of
event times. The argument of the logarithm is independent of $\lbrace
t_-\rbrace$, and hence this integral can be performed trivially. We are left
with exactly the mutual information between $s(t)$ and $\lbrace t_+\rbrace$,
\begin{equation}
	I(s(t),x(t))=I(s(t),\lbrace t_+\rbrace).
\end{equation}
This result shows that the information about $S$ which can be extracted from
$x(t)$ is contained specifically in the timing of $X$ production events. 

The discussion above assumes that individual production and decay steps can be
resolved and that the timing of all events is known exactly, and thus is only
valid if we can observe the continuous trajectory $x(t)$ on all timescales. If
on the other hand we only have a discrete sampling $\mathbf{x}=(x(t_1),
x(t_2),\dots,x(t_N))$, then the degradation reactions of $X$ increase our
uncertainty about $S$. If we observe that the number of $X$ molecules has
changed by $n=x(t_i)-x(t_{i-1})$ over the interval $[t_{i-1},t_i]$, then we can
conclude that the number of production events in the corresponding interval
minus the number of degradation events must equal $n$. However, we do not know
exactly how many production and decay events have taken place, and therefore the
accuracy with which we can estimate the input trajectory $s(t)$ is reduced.

\subsubsection{Motif III} \label{sec:res_traj_3}

For this motif the gain-to-noise ratio is 
\begin{equation}
	\frac{g^2(\omega)}{N(\omega)}=\frac{\rho}{2\langle S\rangle},
	\label{eq:gnr_motif3}
\end{equation}
independent of $\omega$. This motif is therefore able to transmit signals at all
frequencies equally well. However, we note that both the gain and noise,
\begin{eqnarray}
	g^2(\omega)&=&\frac{\rho^2}{\omega^2+\mu^2} \\
	N(\omega)&=&\frac{2\rho\langle S\rangle}{\omega^2+\mu^2}
\end{eqnarray}
decrease at frequencies larger than $\mu$. Both the input signal and the
intrinsic noise in $X$ are effectively integrated over the lifetime of $X$
molecules, $1/\mu$, and so are attenuated at high frequencies. It should be
noted that while information can be reliably encoded at high frequencies in the
signal $x(t)$, the power associated with these variations,
$g^2(\omega)P_{ss}(\omega)$, may be small. Therefore, if this signal is taken
as the input to another downstream process, these signals may be difficult to
decode. For example, intrinsic noise in the detection of $X$ by the downstream
network may overwhelm the small amplitude signal at high frequencies 
\cite{Tostevin09}.

The coherence for this motif is
\begin{equation} \label{eq:coh_motif3}
	\phi(\omega)=\frac{\rho\lambda}{\omega^2+\lambda(\lambda+\rho)},
\end{equation}
showing that the information content of the output signal decreases at high
frequencies. This is because the coherence depends on the input power spectrum
$P_{ss}(\omega)$, which itself scales as $\omega^{-2}$ for $\omega\gg\lambda$;
the information content of the input signal is itself reduced at high
frequencies. The mutual information rate (in nats per unit time) for this motif
is 
\begin{equation} \label{eq:R_motif3}
	R(S,X)=\frac{\lambda}{2}\left[\sqrt{1+\rho/\lambda}-1\right].
\end{equation}
As we saw for motif II, both the coherence and information rate are independent
of $\mu$, the decay rate of $X$. As discussed above, since decay events of $X$
occur independently of the input signal $S$, the information we can gain about
the input signal is the information encoded in the production of $X$. 

We may therefore be tempted to conclude that the decay events of $X$ contain no
information about the input signal $s(t)$. This is not entirely true. The
information encoded in the decay events of $X$ can be quantified by considering
a modified network combining motifs II and III as follows:
\begin{equation} \begin{array}{c}
	\emptyset \xrightarrow{\kappa} S,\ 
	S\xrightarrow{\lambda} \emptyset, \\
	S\xrightarrow{\rho} S+X,\ 
	X\xrightarrow{\mu}Y,\ 
	Y\xrightarrow{\mu_Y}\emptyset. \end{array}
	\label{eq:motif3deg}
\end{equation}
As we saw in the previous section, in motif II information about the input
signal is encoded in the timing of $X$ production reactions. Similarly, in the
set of reactions in Eq. \ref{eq:motif3deg} information about $S$ will be encoded
in the timing of $Y$ production reactions. Since the reactions producing $Y$
correspond to decays of $X$, the mutual information between $s(t)$ and $y(t)$
will also be the mutual information between $s(t)$ and the decays of $X$. In the
limit $\mu\rightarrow\infty$, this set of reactions reduces to motif III: as $X$
molecules decay immediately the two reactions $S\xrightarrow{\rho}S+X$ and
$X\xrightarrow{\mu}Y$ are effectively combined into $S\xrightarrow{\rho}S+Y$.
However, for finite $\mu$ we find that the mutual information rate between $S$ 
and $Y$,
\begin{equation} \label{eq:R_3_decay}
	R(S,Y)=\frac{\lambda}{2}
		\left[\sqrt{1+2\alpha\sqrt{1+\beta}+\alpha^2}-(1+\alpha)\right],
\end{equation}
where $\alpha=\mu/\lambda$ and $\beta=\rho/\lambda$, is reduced compared to Eq.
\ref{eq:R_motif3}. Additionally, the gain-to-noise ratio and coherence,
\begin{eqnarray}
	\frac{g^2(\omega)}{N(\omega)}&=&\frac{\rho}{2\langle S\rangle}
		\frac{\mu^2}{\omega^2+\mu^2},\\
	\phi(\omega)&=&\frac{\mu^2\rho\lambda}
		{\mu^2\rho\lambda+(\omega^2+\lambda^2)(\omega^2+\mu^2)},
\end{eqnarray}
are reduced at all frequencies compared to Eq. \ref{eq:gnr_motif3} and Eq.
\ref{eq:coh_motif3}. Therefore we can see that the decay events of $X$ do
provide some information about the trajectory $s(t)$, but less than can be
obtained from observing the production of $X$. Since decays of $X$ take place
independently of $S$, from observing the decay of $X$ we can only estimate,
based on the lifetime of $X$, a distribution of times in the past when an $S$
molecule was present. However, observing the production of an $X$ molecule at a
given time $t$ tells us directly that at least one $S$ molecule is present. We
conclude that in the case where we can observe the entire trajectory of $X$,
decay events provide us with no {\em additional} information which we could not
already extract from the production events contained in the trajectory. 

\subsubsection{Comparison of different motifs} \label{sec:comp}

It is important to realise that in motifs I and II considered above the dynamics
of the detection reaction affects the statistics of the signal. The
gain-to-noise ratios for these motifs are therefore not intrinsic network
properties, but also depend on the input process. This can be seen from the
appearance of the degradation rate for the input species, $\lambda$, in the
expressions for the gain-to-noise ratio. In general, therefore, the ensemble of
input signals will be different for each motif, making it problematic to
directly compare the different information rates. However, we can make a useful
comparison in cases in which the input signals have the same form.

First we shall examine motifs II and III, and consider the special case where
the statistics of $s(t)$ are the same for both reactions. To achieve this we
choose in motif II $\lambda=0$, and in motif III $\lambda=\rho$. Both motifs can
then be described by the same macroscopic evolution equations,
\begin{subequations} \label{eq:comp_23} \begin{eqnarray}
	\diff{s}{t}&=&\kappa-\rho s(t) +\eta_s(t) \\
	\diff{x}{t}&=&\rho s(t)-\mu x(t) +\eta_x(t). 
\end{eqnarray} \end{subequations}
However, the noise correlations will be different in the two cases: for motif II
we have $\langle\eta_s(t)\eta_x(t')\rangle=-\rho\langle S\rangle\delta(t-t')$,
while for motif III $\langle\eta_s(t)\eta_x(t')\rangle=0$. Calculating the
coherence of these two systems gives
\begin{eqnarray}
	\phi^{\rm II}(\omega)&=&\frac{1}{4} \label{eq:comp_2_coherence}\\
	\phi^{\rm III}(\omega)&=&\frac{1}{2+\omega^2/\rho^2}.
		\label{eq:comp_3_coherence}
\end{eqnarray}
Since the input power spectra $P_{ss}(\omega)$ are the same for these two
networks, we can use the coherence directly as a comparison of signalling
performance for signals of different frequencies. We see that at low
frequencies, $\omega<\sqrt{2}\rho$, motif III provides more information, while
for high frequencies, $\omega>\sqrt{2}\rho$, motif II allows for more reliable
signal transmission. 

\begin{figure}
	\includegraphics{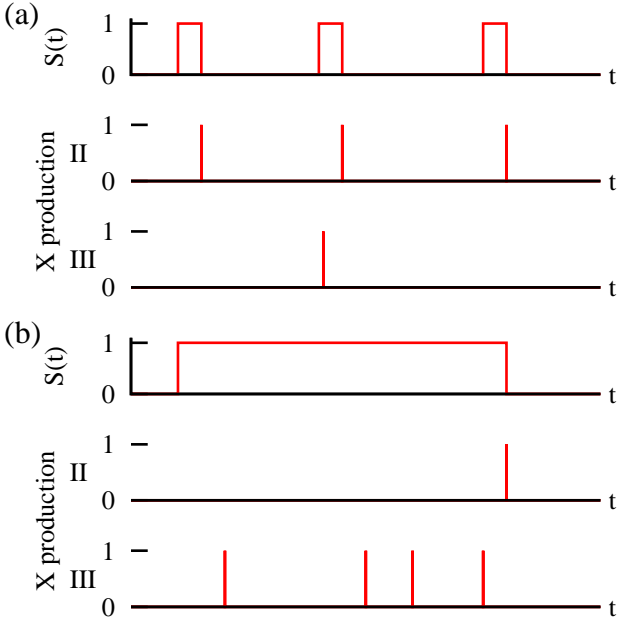}
	\caption{Comparison of $X$ production in motifs II and III for
		$S$ molecules with different lifetimes. In each panel the upper plot
		shows an example of the birth and death of $S$ molecules. The second
		shows the corresponding timing of $X$ production in motif II. The lower
		trace shows an example of $X$ production events in motif III.  (a) For
		some short-lived $S$ molecules no $X$ will be produced in motif III. We
		therefore lose information about these signals, in contrast to motif II
		for which all $S$ molecules are detected.  (b) For long-lived $S$
		molecules motif III allows for a more accurate estimate of the
		production time, since more than one $X$ molecule can be produced. This
		increases the total information about $s(t)$ compared to motif II.} 
	\label{fig:II-III}
\end{figure}

To understand these differing responses of $X$ to signals at different
timescales we consider the information that can be extracted from the trace 
$x(t)$ about $S$ molecules with different lifetimes, as depicted in Fig.
\ref{fig:II-III}. The spontaneous decay reaction we have assumed for $S$ means
that the lifetimes of different $S$ molecules will be exponentially distributed
with mean $1/\rho$. Recall also that the information about the signal $s(t)$ is
contained in the timing of the production reactions for $X$. Since we have taken
$\lambda=0$ in motif II, we know that the decay of every $S$ molecule
corresponds to the production of an $X$. We therefore obtain the same amount of
information about the production and decay of each $S$ molecule, regardless of
its lifetime. We learn exactly the time at which it decays. In addition we can
estimate, based on the typical lifetime of an $S$ molecule, the time at which
the molecule was produced. However, for $S$ molecules with a lifetime much
longer or shorter than the average this estimate will be inaccurate. 

In motif III, on the other hand, $X$ molecules will be produced at an average
rate $\rho$ for each $S$ molecule present. Thus for $S$ molecules which decay
extremely rapidly with a short lifetime $\tau\ll1/\rho$, the probability of
producing an $X$ molecule will be small. We effectively do not detect these $S$
molecules at all, and hence gain no information about their contribution to the
trajectory $s(t)$. This is in marked contrast to motif II, for which we know
that we will detect all $S$ molecules. On the other hand, for $S$ molecules with
a much longer lifetime $\tau\gg1/\rho$, more than one $X$ molecule will be
produced on average. From this sequence of production events we can estimate
both the production and decay times of the corresponding $S$. Importantly, while
we do not get as much information about the decay time as we would in motif II,
our estimate of the production time will be more accurate than motif II allows.
The total information gained about both the production and decay times may
therefore be higher. On average motif III provides more information than motif
II about $S$ molecules with a longer than average lifetime, but less information
about $S$ molecules with a short lifetime. At a more macroscopic level, motif
III effectively amplifies slowly-varying signals, producing more than one $X$
molecule for each $S$, but averages over extremely rapid changes in $S$,
producing less than one $X$ per $S$. Motif II transmits all signals with a
similar amplitude, regardless of the timescale of the input.

Another way to construct a system in which the reactions under comparison do not
affect the input signal is to place the input signal upstream of $S$, and to use
this uncorrelated process to drive the reactions of interest. We therefore add
to each motif in Table \ref{tab:motifs} a signal $Q$, which drives the
production of $S$ via the reaction $Q\torate{\alpha}Q+S$. This choice ensures
that the input signal, now $q(t)$, is uncorrelated from the noise within the
downstream signalling network, and hence that the input power spectrum
$P_{qq}(\omega)$ is unchanged by the network dynamics. Then the fidelity of
signal transmission between $q(t)$ and the output $x(t)$ can be quantified by
the gain-to-noise ratio, $g^2(\omega)/N(\omega)$. The gain-to-noise ratios of
these modified reaction motifs are shown in Table \ref{tab:gnr}.

\begin{table}
\begin{tabular}{cc}
Motif & $g^2(\omega)/N(\omega)$
\\[3pt]\hline\\[-10pt]
	(I)
& $\frac{\alpha\rho}{2\langle Q\rangle}\left[
		\frac{\lambda}{\omega^2+\lambda(\lambda+\rho)}
	\right]$
\\[5pt]\hline\\[-10pt]
	(Ia)
& $\frac{\alpha\rho}{2\langle Q\rangle}\left[
		\frac{\lambda(\lambda+\rho+\mu)}
		{(\lambda+\mu)\omega^2+\lambda(\lambda+\rho)(\lambda+\rho+\mu)}
	\right]$
\\[5pt]\hline\\[-10pt]
	(II) 
& $\frac{\alpha\rho}{2\langle Q\rangle}\left[
		\frac{\lambda+\rho}{\omega^2+(\lambda+\rho)^2}
	\right]$
\\[5pt]\hline\\[-10pt]
	(III)
& $\frac{\alpha\rho}{2\langle Q\rangle}\left[
		\frac{\lambda}{\omega^2+\lambda(\lambda+\rho)}
	\right]$
\\[5pt]\hline
\end{tabular}
	\caption{The gain-to-noise ratio of the reaction motifs in Table 
		\ref{tab:motifs} when driven by an upstream signal $Q$ via 
		$Q\torate{\alpha}Q+S$. The row labelled Ia corresponds to motif I but with 
		the additional reaction $X\torate{\lambda}\emptyset$.
	}
	\label{tab:gnr}
\end{table}

The expressions in Table \ref{tab:gnr} have a number of interesting features.
Firstly, we note that for equal reaction rates the gain-to-noise ratios of
motifs I and III are identical. In both motifs I and III we can detect the same
$S$ molecule many times, either by repeated switching between the $S$ and $X$
forms or because one $S$ can stimulate the production of several $X$ molecules.
For these two motifs, with identical rates, one can straightforwardly show that
the distribution of the number of times a given molecule converts from $S$ to
$X$ and back in motif I is identical to the distribution of the number of $X$
molecules produced from a single $S$ molecule in motif III. Thus the strength of
this ``interference'' between repeated detections of the same $S$ molecule is
the same in the two motifs; from the point of view of information transmission,
the two reactions are then equivalent.

In the limit $\omega\to0$, all three expressions tend to the same value. All
three motifs perform equally well for the transmission of slowly-varying
signals. For $\omega>0$ the gain-to-noise ratio for motif II becomes larger than
that of either motif I or III, showing that motif II is able to transmit more
information. Comparing motifs II and III, at low frequencies the gain is lower
in motif II as we found above when considering signal transmission from $S$ to
$X$. However, the noise is also lower in motif II at all frequencies, because
intrinsic noise in the production of $S$ does not propagate to $X$ (see Appendix
\ref{sec:app} for further details). Comparing motifs I and II, the situation is
more complex. At low frequencies the gain is smaller in motif II; however, the
noise is once again also less in motif II, making these motifs perform equally 
well for slowly-varying signals. At frequencies $\omega^2>\mu(\lambda+\rho/2)$
the gain is larger for motif II, as the switching reaction $S\tofrom X$ is less
able to track rapid changes in $S$. At intermediate frequencies it is possible
for motif I to have a lower noise power than motif II. However, in this regime
the gain in motif I is significantly smaller than that of motif II, and so motif
II is still able to transmit more information at these frequencies.

\begin{figure}
	\includegraphics{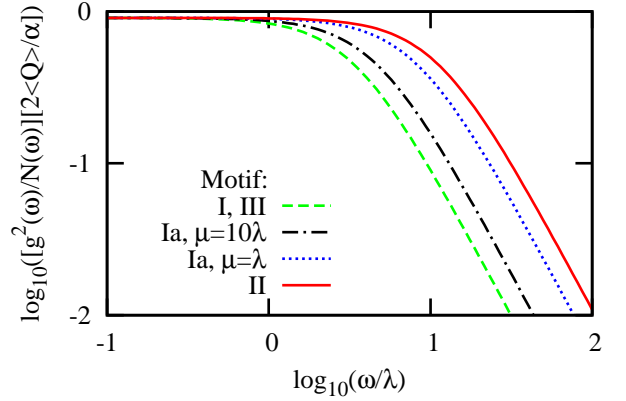}
	\caption{Gain-to-noise ratio as a function of frequency
		between $q(t)$ and $x(t)$ for different motifs. The gain-to-noise ratio
		is highest for motif II, and lowest for motifs I and III. As $\mu$ is
		varied in motif Ia, the gain-to-noise ratio interpolates between these
		two extremes. In each case $\rho=10\lambda$ was chosen.}
	\label{fig:gnr_comp}
\end{figure}

We can also consider a modified version of motif I in which $X$ molecules can
also degrade spontaneously via the same reaction as $S$, i.e.
$X\xrightarrow{\lambda}\emptyset$. We denote this set of reactions by motif Ia,
and the corresponding gain-to-noise ratio is shown in Table \ref{tab:gnr}. The
probability of an $X$ molecule switching back to the $S$ form depends on the
relative rates of the two possible decay reactions for $X$,
$X\xrightarrow{\mu}S$ and $X\xrightarrow{\lambda}\emptyset$, and is given by
$\mu/(\mu+\lambda)$. In the limit $\mu\to0$, this motif reduces to motif II: $X$
molecules never switch back to $S$ but instead always decay. In the limit
$\mu\to\infty$, $X$ molecules always return to the $S$ state, and hence we
recover motif I. By varying $\mu$ we therefore change the number of times the same
molecule switches between the $S$ and $X$ states. We would therefore expect the
gain-to-noise ratio for this motif to increase with decreasing $\mu$. We can see
from Table \ref{tab:gnr} that for any finite non-zero value of $\mu$ the
gain-to-noise of motif Ia is indeed larger at all frequencies than that of motif
I, but not as large as that of motif II. As $\mu$ is decreased the gain-to-noise
ratio for motif Ia interpolates smoothly between those of motifs I and II, as
shown in Fig. \ref{fig:gnr_comp}. 

Finally, we examine motifs II and III with the combination of both an upstream
signal $Q$ and the parameter choices of Eq. \ref{eq:comp_23}: $\lambda=0$ in 
motif II and $\lambda=\rho$ in motif III. Specifically, we consider two networks
described by 
\begin{subequations} \label{eq:comp_23_Q} \begin{eqnarray}
	\diff{s}{t}&=&\alpha q(t)-\rho s(t)+\eta_s(t) \\
	\diff{x}{t}&=&\rho s(t)-\mu x(t)+\eta_x(t),
\end{eqnarray} \end{subequations}
where as before $\langle\eta_s(t)\eta_x(t')\rangle=-\rho\langle S\rangle
\delta(t-t')$ or $\langle\eta_s(t)\eta_x(t')\rangle=0$ for motifs II and III
respectively. For these parameter choices the power spectra of $Q$ and $S$ are
the same in both networks. 

\begin{figure}
	\includegraphics{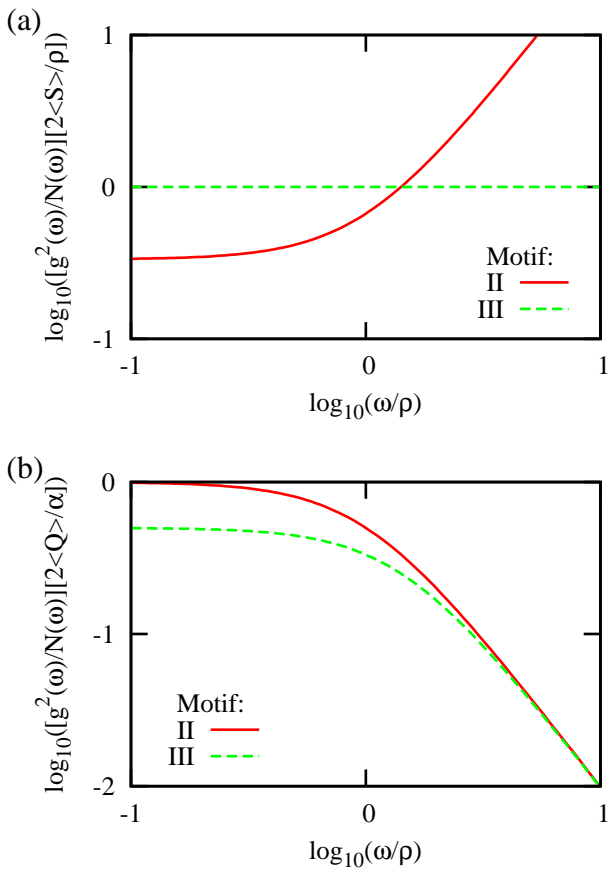}
	\caption{Comparison of the performance of motifs II and III
		for the transmission of different input signals.  (a) The gain-to-noise
		for the transmission of $s(t)$ to $x(t)$ for the basic motifs with
		parameters chosen according to Eq. \ref{eq:comp_23}: $\lambda=0$ in
		motif II and $\lambda=\rho$ in motif III.  (b) Gain-to-noise ratio for
		the transmission of $q(t)$ to $x(t)$, as in Eq.  \ref{eq:comp_23_Q}
		with the same parameter choice.}
	\label{fig:23_comp}
\end{figure}

The gain-to-noise ratios between the input signal $q(t)$ and output $x(t)$ for
the two networks are 
\begin{eqnarray}
	\left[\frac{g^2(\omega)}{N(\omega)}\right]^{\rm II}&=& \label{eq:comp_2_Q}
		\frac{\alpha}{2\langle Q\rangle}\left[\frac{1}{(\omega/\rho)^2+1}\right] \\
	\left[\frac{g^2(\omega)}{N(\omega)}\right]^{\rm III}&=& \label{eq:comp_3_Q}
		\frac{\alpha}{2\langle Q\rangle}\left[\frac{1}{(\omega/\rho)^2+2}\right].
\end{eqnarray}
Interestingly we now find that motif II can transmit low-frequency signals
($\omega\ll\rho$) more reliably than motif III, while for high frequency signals
($\omega\gg\rho$) the two reactions perform equally well. This is in contrast to
Eqs \ref{eq:comp_2_coherence} and \ref{eq:comp_3_coherence}, which show that
motif III is able to more reliably transmit information about $s(t)$ at low
frequencies while motif II is more reliable at high frequencies (see Fig.
\ref{fig:23_comp}). 

To understand how the fidelity of signal transmission changes for different
input signals we must consider the different sources and propagation of noise in
these networks (see Appendix \ref{sec:app} for more details). If the input
signal is taken to be $s(t)$ then any fluctuations in $S$, regardless of their
origin, contribute to the input signal. However, when we are concerned with the
transmission of $q(t)$, fluctuations in $s(t)$ that are uncorrelated from $q(t)$
are considered noise. We wish to transmit only those changes in $S$ that are
caused by $Q$. For the two motifs considered here the mean response to an input
signal $q(t)$, measured by the gain between $q(t)$ and $x(t)$, is the same.
However, the propagation of intrinsic noise in $S$ differs between the two
motifs. We saw previously that motif III is able to amplify fluctuations in $S$
at low frequencies ($\omega\ll\rho$). The fluctuations which made up the signal
$s(t)$ in the network described by Eq. \ref{eq:comp_23} now correspond in Eq.
\ref{eq:comp_23_Q} to intrinsic noise, independent of $q(t)$, in the production
and decay of $S$. If the input signal of interest is $s(t)$ then amplification
of these fluctuations is beneficial, as we saw in Fig. \ref{fig:II-III}, as it
allows for better resolution of different $s(t)$ signals. However, if the input
signal is $q(t)$, then amplifying intrinsic fluctuations in $S$, which are
uncorrelated from $Q$, provides no information about $q(t)$; indeed this noise
obscures the desired signal. Motif II, meanwhile, does not suffer from this
problem as it cannot amplify intrinsic fluctuations in $S$; indeed, noise in the
production of $S$ is not propagated to $X$. Consequently at low frequencies the
noise in the transmission of $q(t)$ is larger in motif III than in motif II,
reducing the fidelity of transmission of these signals. For high frequency
($\omega\gg\rho$) signals we saw previously that motif III is unable to track
changes in $S$, and hence the transmitted noise decreases. The total noise power
for motif III is then dominated by intrinsic noise in the production and decay
of $X$, which is the same as that found in motif II. Therefore, the
gain-to-noise ratios of the two motifs become the same for high-frequency
signals.

\section{Discussion}

In this paper we have considered the transmission of time-varying signals
through elementary biochemical reactions. We have considered the transmission
of both instantaneous signals and of complete trajectories. Our results show
that these reactions can have radically different information characteristics
for these two types of signals. Most strikingly, for the irreversible
modification reaction in motif II the instantaneous information is zero, yet for
trajectories we find an extremely large mutual information rate. Additionally
for reaction motif III the mutual information rate is independent of the decay
rate of the output component, but the instantaneous mutual information does
depend on this parameter. These two information quantities therefore measure
different aspects of the signalling behaviour of networks, both of which may be
important in different biochemical systems.

The reliability with which a network can transmit a particular frequency
component of the input signal trajectory is determined by the gain-to-noise
ratio of the network as a function of frequency. For systems that obey the
spectral addition rule \cite{TanaseNicola06}, that is those for which downstream
reactions do not affect the input signal, the gain-to-noise ratio is an
intrinsic property of the processing network. For networks that do not obey the
spectral addition rule the gain-to-noise ratio will be dependent on the
statistics of the input signal. The mutual information between input and output
signals, which quantifies the information which can be transmitted about a
particular input ensemble, also depends on the particular choice of the input
signal. Thus when comparing the mutual information in different motifs, or the
gain-to-noise ratio of motifs for which the statistics of the input signal are
affected by the network, care should be taken to ensure that the input
distributions of the different networks are the same.

Recently Endres and Wingreen showed \cite{Endres08} that an absorbing detector
is able to more accurately measure a steady-state ligand concentration than
either a detector that allows for passive observation of the local concentration
or a detector to which a ligand can bind reversibly. These different situations
are analogous to our motifs II, III and I respectively. It has also been
observed previously that the irreversible conversion reaction $S\to X$ reduces
the propagation of noise through networks \cite{TanaseNicola06, LevineE07}. Our
results in section \ref{sec:comp} show that the optimal reaction to transmit
time-varying signals depends on the input signal that the network is attempting 
to transmit. If the signal is the concentration of ligand itself, as in
Fig. \ref{fig:23_comp}(a), then an absorbing detector is beneficial only for
rapidly-varying ligand signals, since this reaction does not allow for
amplification of slowly-varying signals. To detect low-frequency changes in the
ligand signal it is better to respond via a reaction that does not consume the
ligand molecule. On the other hand, if the ligand is itself a reporter for
another upstream process then the production reaction of motif III amplifies
intrinsic noise in the signalling network at low frequencies, obscuring the
signal of interest. In this case, as we saw in Figs. \ref{fig:gnr_comp} and
\ref{fig:23_comp}(b), the use of an absorbing detector is always preferable.

The qualitative difference in the relative performance of motifs II and III for
different input signals, shown in Fig. \ref{fig:23_comp}, has important
consequences for the design of signalling cascades. Intuitively we might expect
that selecting the reaction for each level of the cascade which provides the
most information about the dynamics at the previous level would optimise
information transmission for the entire cascade. However, Fig. \ref{fig:23_comp}
shows that this is not the case. In fact, choosing the reaction which optimises
the reliability of signalling at one downstream step can {\em reduce} the
overall information transmitted through the cascade if more intrinsic noise is
propagated. Our results suggest that even in the absence of feedback it is vital
to consider the transmission properties of signalling cascades as a whole,
rather than isolating individual reaction steps. Naturally-occurring signalling
cascades may consist of a number of steps which are individually sub-optimal in
terms of one component tracking fluctuations in another, but which together
optimise the transmission of the signal of interest while minimising the impact
of noise within the network. 

We have observed that the information rates for reaction motifs II and III are
independent of the decay rate of the output component $X$. This decay rate sets
the relaxation time for intrinsic fluctuations in $X$ and is often considered to
set the timescale on which $X$ is able to respond to signals. However, in terms
of information transmission this is not the case. Instead of the dissipative
timescale, the ability of $X$ to respond to changes in the input signal is
determined by the rate of {\em production} events. Importantly, the mutual
information is independent of the decay rate of only the {\em output} component
of a signalling pathway. Information transmission does depend on the relaxation
rate of intermediate signalling components. Thus if $X$ is taken as the input to
another process then the overall information rate will depend on its relaxation
rate, as we saw when considering the decay of $X$ to $Y$ in Section
\ref{sec:res_traj_3}. More generally, our results indicate that signals in
biochemical networks can be encoded in the timing of specific reaction events.
To understand whether cells can exploit this information it will be important to
investigate situations in which different encoding strategies are employed {\em
in vivo}, and to understand the ability of cells to decode this information.

The mutual information between trajectories, as we have calculated here, is a
measure of how reliably signals can be transduced through networks. However, it
is known that many biochemical networks also perform other signal-processing
functions such as filtering high-frequency \cite{Bennett08} or low-frequency
input signals \cite{Segall86, Mettetal08}. While these response characteristics
of networks appear in the gain-to-noise ratio, as we have shown here and
previously \cite{Tostevin09, deRonde09a, deRonde09b}, the mutual information
does not distinguish between the properties of the input signal that the cell
wishes to respond to and those with which the cell is not concerned. In these
cases a more biologically-meaningful measure of the performance of the network
would be the mutual information between the properties of interest in the input
signal and the output trajectory. For example, if a cell wishes to decode the
frequency of an oscillating input signal into the amplitude of a messenger
signal, but is not concerned with the amplitude or phase, a more appropriate
measure of signalling would be the mutual information between the input signal
frequency and the output signal amplitude. The appropriate input and output
signals must be considered on a case-by-case basis, and relies on our
understanding of the biological function of the particular system being
considered. We hope to address these issues in more detail in future work.

Throughout this paper we have calculated information transmission for a Gaussian
model of the network of interest. As discussed in Section \ref{sec:model}, such
a model is also able to provide a lower bound on the channel capacity for
non-Gaussian systems, provided that the Gaussian model is chosen appropriately.
Even for the linear systems considered in Section \ref{sec:results}, the
calculated results are strictly only lower bounds on the channel capacity of
real biochemical systems. It is known that for many such systems, particularly
for copy numbers of order a few hundred molecules which are often found in
signalling networks, the approximation of small Gaussian noise is very accurate
\cite{Ziv07, TanaseNicola06, Bruggeman09, deRonde09a}. However, in most cases it
is not clear what the typical input distributions of biochemical networks in
natural environments are. It is therefore difficult to quantify the impact of
assuming a Gaussian input distribution in this analysis. We hope that future
experiments will clarify the typical distributions of environmental stimuli to
which cells are exposed, and which are crucial in determining how information is
propagated through networks.

\acknowledgments
We thank Wiet de Ronde for extensive discussions, and Debasish Chaudhuri for a
critical reading of the manuscript. This work is part of the research program of
the ``Stichting voor Fundamenteel Onderzoek der Materie (FOM)'', which is
financially supported by the ``Nederlandse organisatie voor Wetenschappelijk
Onderzoek (NWO)''.

\appendix
\section{Comparison of signal and noise partitioning for motifs II and III} 
\label{sec:app}

Here we consider in more detail the partitioning of the output power
$P_{xx}(\omega)$ into signal and noise contributions for transmission from $Q$
to $X$ and from $S$ to $X$. From the Langevin equations in Eq. 
\ref{eq:comp_23_Q}, where we have chosen $\lambda=0$ in motif II and 
$\lambda=\rho$ in motif III, the output power spectrum of both motifs can be 
written as 
\begin{multline}
	P_{xx}(\omega)=
		\frac{\rho^2}{\omega^2+\mu^2}\frac{\alpha^2}{\omega^2+\rho^2}P_{qq}(\omega)
	+ \frac{\rho^2}{\omega^2+\mu^2}\frac{\Gamma_{ss}}{\omega^2+\rho^2}\\
	+ \frac{\Gamma_{xx}}{\omega^2+\mu^2}
	+ \frac{2\rho^2}{\omega^2+\mu^2}\frac{\Gamma_{sx}}{\omega^2+\rho^2}, 
	\label{eq:app_Pxx}
\end{multline}
where the $\Gamma_{\alpha\beta}$ factors describe the various noise strengths
and correlations and are defined by $\langle\eta_\alpha(t)\eta_\beta(t')\rangle=
\Gamma_{\alpha\beta}\delta(t-t')$. We note that while we do not specify a
particular process or power spectrum for $q(t)$, we have assumed that the noise
in $S$ is uncorrelated from $Q$: $\langle q(t)\eta_s(t')\rangle=0$ for all $t$
and $t'$. The first term in Eq. \ref{eq:app_Pxx} describes the influence of
$q(t)$ on $x(t)$ in the absence of noise, and characterises the mean response of
$x(t)$ to changes in $q(t)$. The two prefactors to $P_{qq}(\omega)$ represent
the effective response functions at the two levels on the cascade, the
transmission of $Q$ to $S$ and of $S$ to $X$. These transfer functions show that
high-frequency signals are attenuated at each step. The second term in Eq.
\ref{eq:app_Pxx} includes intrinsic noise in the production and decay of $S$
molecules. The third term similarly describes intrinsic noise in the production
and decay of $X$. For the reactions and parameters we are considering, these
three terms are the same for both motifs, with $\Gamma_{ss}=\Gamma_{xx}=
2\alpha\langle Q\rangle$. The final term contains corrections to the two
previous noise terms due to correlations between $\eta_s(t)$ and $\eta_x(t)$.
For motif III we have $\Gamma_{sx}=0$. However, for motif II
$\Gamma_{sx}=-\alpha\langle Q\rangle$, and this term precisely cancels the
intrinsic noise in $S$. This latter noise, the second term in Eq.
\ref{eq:app_Pxx}, contains contributions from the production reactions of $S$
and from the decay reactions. Since this decay occurs via the reaction $S\to X$
in motif II, these events are also included in the term describing intrinsic
noise in the production of $X$, the third term in Eq. \ref{eq:app_Pxx}. The
negative cross-correlation term therefore eliminates the double-counting of
these events. Furthermore, the cross-correlation term also removes fluctuations
in $X$ due to noise in the production of $S$. This noise does not propagate to
$X$ in motif II, because in the regime of small fluctuations around steady state 
the spontaneous production and decay of $S$ molecules are two independent 
Poisson processes; hence noise in the production of $S$ is uncorrelated from the 
production of $X$. This is in contrast to motif III, for which production of $X$
molecules can occur whenever an $S$ is present, and hence the signal $x(t)$
depends on the noise in both the production and decay of $S$. Ultimately, for
the two motifs we have 
\begin{equation}
	P_{xx}^{\rm II}(\omega)=
		\frac{\rho^2}{\omega^2+\mu^2}\frac{\alpha^2}{\omega^2+\rho^2}P_{qq}(\omega)
		+\frac{\Gamma_{xx}}{\omega^2+\mu^2} \label{eq:Pxx_2} 
\end{equation}
\begin{multline}
	P_{xx}^{\rm III}(\omega)=
		\frac{\rho^2}{\omega^2+\mu^2}\frac{\alpha^2}{\omega^2+\rho^2}P_{qq}(\omega)
		\\+\frac{\rho^2}{\omega^2+\mu^2}\frac{\Gamma_{ss}}
			{\omega^2+\rho^2}+\frac{\Gamma_{xx}}{\omega^2+\mu^2}. \label{eq:Pxx_3}
\end{multline}

If we consider the transmission of $Q$ to $X$, then only fluctuations in $x(t)$
which are correlated with $q(t)$ should contribute to the signal component of
the output power. It also follows from the definition in Eq. \ref{eq:signal_def}
that $\Sigma_{q\to x}(\omega)$ is the same for both motifs,
\begin{equation}
	\Sigma_{q\to x}(\omega)=\frac{\rho^2}{\omega^2+\mu^2}
		\frac{\alpha^2}{\omega^2+\rho^2}P_{qq}(\omega).
\end{equation}
The remaining terms in Eq. \ref{eq:app_Pxx} form the noise power spectrum. For
motif II the noise in the output signal is 
\begin{equation}
	N^{\rm II}_{q\to x}(\omega)=\frac{\Gamma_{xx}}{\omega^2+\mu^2},
\end{equation}
while for motif III we have 
\begin{equation}
	N^{\rm III}_{q\to x}(\omega)=\frac{\rho^2}{\omega^2+\mu^2}
		\frac{\Gamma_{ss}}{\omega^2+\rho^2} + \frac{\Gamma_{xx}}{\omega^2+\mu^2}.
\end{equation}
We can therefore see that the total noise is smaller in motif II than in motif
III because, as  noted above, noise in the production of $S$ does not propagate 
to $X$. However, this difference becomes negligible at high frequencies
$\omega\gg\rho$ because at these frequencies motif III effectively averages over
rapid fluctuations in $S$ and hence the effect of these fluctuations on $X$
diminishes.

Now suppose that we take $s(t)$, rather than $q(t)$, to be the input signal to
the network. Then the signal component of the output power should include those
terms for which $X$ is correlated with $S$. For motif III we can
straightforwardly see that the second term in Eq. \ref{eq:app_Pxx}, which
describes fluctuations in $x(t)$ due to intrinsic noise in $S$, should
contribute to the signal power. The intrinsic fluctuations in $X$ remain
uncorrelated from $S$, and hence are still considered noise. We therefore have
\begin{eqnarray}
	\Sigma^{\rm III}_{s\to x}(\omega)&=&
		\frac{\rho^2}{\omega^2+\mu^2}
			\left[\frac{\alpha^2}{\omega^2+\rho^2}P_{qq}(\omega)+
				\frac{\Gamma_{ss}}{\omega^2+\rho^2}\right] \label{eq:app_sx_III_signal}
	\\ N^{\rm III}_{s\to x}(\omega)&=&\frac{\Gamma_{xx}}{\omega^2+\mu^2}.
\end{eqnarray}
We can also identify the bracketed terms in Eq. \ref{eq:app_sx_III_signal} as
the power spectrum of $S$. If we consider motif II, while $P_{xx}(\omega)$ does
not depend on $\eta_s(t)$ explicitly we must recall that noise in the production
of $X$ molecules corresponds to noise in the decay of $S$, and hence contributes
to the signal power. Applying the definition of the signal power, Eq.
\ref{eq:signal_def}, leads to the following result: 
\begin{multline}
	\Sigma^{\rm II}_{s\to x}(\omega)= \label{eq:app_sx_II_signal}
		\frac{\rho^2}{\omega^2+\mu^2}\frac{\alpha^2}{\omega^2+\rho^2}P_{qq}(\omega)
	\\+ \frac{\Gamma_{sx}}{\omega^2+\mu^2}
			\frac{\Gamma_{sx}}{\left[\alpha^2P_{qq}(\omega)+\Gamma_{ss}\right]}
\end{multline}
\begin{equation}
	N^{\rm II}_{s\to x}(\omega)=\frac{\Gamma_{xx}}{\omega^2+\mu^2}
		-\frac{\Gamma_{sx}}{\omega^2+\mu^2}
			\frac{\Gamma_{sx}}{\left[\alpha^2P_{qq}(\omega)+\Gamma_{ss}\right]}.
\end{equation}
The noise in motif II is again lower than in motif III at all frequencies. In
this case, this difference arises because in motif II noise in the production of
$X$ is correlated with $s(t)$. To study the differences between the signal
powers of the two motifs we consider 
\begin{multline}
	\Delta\Sigma(\omega)=
		\Sigma^{\rm II}_{s\to x}(\omega)-\Sigma^{\rm III}_{s\to x}(\omega)\\
		= \frac{\Gamma_{ss}}{\omega^2+\mu^2}\left[
		\frac{\Gamma_{ss}}{4(\alpha^2P_{qq}(\omega)+\Gamma_{ss})}
		-\frac{\rho^2}{\omega^2+\rho^2} \right], \label{eq:app_sx_diff_signal}
\end{multline}
where we have used the facts that $\Gamma_{ss}$ is the same for both motifs and 
that for motif II $\Gamma_{sx}=-\Gamma_{ss}/2$. We can see that at low
frequencies $\omega\ll\rho$, $\Delta\Sigma(\omega)$ is negative, showing that
the signal power is larger in motif III than in motif II. This reflects
amplification of low-frequency noise in $s(t)$ by motif III. At high frequencies
motif III is no longer able to amplify noise, and instead at the level of $X$
averages over these fluctuations. In this regime $\Delta\Sigma(\omega)$ becomes
positive, showing that the signal power is larger in motif II than in motif III.
This transition results in the cross-over observed in Fig. \ref{fig:23_comp}(a) 
between the different frequency regimes in which motif II or III can more 
reliably transmit signals.

\end{document}